\documentstyle[epsfig,A4,12pt]{article}
\begin{document}
\begin{flushright}

{\bf JINR Preprint~~~~~~\\
E2--95--474~~~~~~~~~\\
Dubna, November 1995}\\
\end{flushright}

\thispagestyle{empty}

 \begin{center}
\vskip 2cm
{\Large\bf { On the universality of the $x$ and $A$ dependence
of the EMC effect and its relation to
parton distributions in nuclei }}
\vskip 1.2cm
{\large  G.I. Smirnov.}\\
\vskip 1. cm
Joint Institute for Nuclear Research, 141980 Dubna, Russia.\\

Current address: CERN, PPE, 1211 Geneva, Switzerland\\
\begin{minipage}[h]{7cm}

email: gsmirnov@cernvm.cern.ch

\end{minipage}

\vskip 1.5 cm

\begin{abstract}
   It is shown that the latest results from the NMC (CERN) and
E665 (Fermilab) groups on  $F_2^A(x)/F_2^D(x)$
obtained in the shadowing region bring
new evidence of the universal $A$  dependence of distortions
made in  a free-nucleon structure function by a nuclear medium.
   The observed universality implies that one can consider
separately hard ($A\leq$ 4) and soft ($A>$ 4)  parton
distribution distortions. Soft distortions, which result in
differencies between the deep-inelastic scattering
cross-sections for nuclei  with masses $A_1$, $A_2 \geq$ 4,
can be explained as a consequence of the nuclear density variation,
independent of $x$ in the range  0.001 $\leq x \leq$ 0.7.

 It is found that nuclear
shadowing begins at $x_{\rm I}$ = 0.0615 $\pm$ 0.0024,
independent of $A$,  which is consistent with models that
allow for three-parton recombination processes.

\end{abstract}

\vspace{1cm}
 Submitted to: {\em Physics Letters B}
\end{center}

\newpage
\baselineskip=1.0\baselineskip
%
%

Particular interest in experimental and theoretical studies
of modifications to the nucleon structure function $F_2(x,Q^2)$
in nuclei, triggered by the discovery
of the EMC effect, is explained by the expectation of finding
a common description for both the  free-nucleon and nuclear
structure functions in the framework of quantum
chromodynamics.

The effects of the distortion of a free-nucleon structure by
a nuclear medium are usually observed as a deviation from
unity of the ratio $r^A(x) \equiv F_2^A(x)/F_2^D(x)$,
where $F_2^A(x)$ and $F_2^D(x)$ are the structure
functions  per nucleon measured in a nucleus of mass $A$
and a deuteron, respectively.

Typical experimental errors in the measurements of the $r^A(x)$
are often of the same order or larger than the values of
distortions. In such a case, the results of a comparison of
distortions obtained at fixed $x$ in different nuclear targets
suffer from large uncertainties.
In particular, this applies to measurements with light nuclei, such
as helium and lithium, which are very important for an understanding
of the $A$ dependence of  the distortions that, by definition
of  $r^A(x)$, should show up for $A \geq$ 3.
On the other hand, the conventional approach, which
 represents the $A$ dependence at fixed $x$ by

\begin{equation}
r^A = C A^{\alpha (x)}~,
\label{slac}
\end{equation}
\noindent does not exploit the conservation of total nucleon
momentum carried by partons. To put it another way, the distortion
of the nucleon structure function by the nuclear medium
at some point $x$ is unjustly considered as
independent of the distortion observed at the adjacent
point $x+ \Delta x$. This has motivated the alternative  approach,
suggested in  Ref.~\cite{sm94}, which  determines the
 $A$ dependence  of distortions after summing them up
over an  interval ($x_1, x_2$).

The analysis of the  data on deep-inelastic
scattering  (DIS) of muons and electrons off nuclear
targets performed
in Ref.~\cite{sm94} demonstrates that the
$A$ dependence of  distortion magnitudes obtained
in each of three regions under study --- namely
shadowing, anti-shadowing and the  EMC effect region ---
follow the same functional form, being different in the
normalizing factor only. This observation gives strong
evidence for the universality of the $x$ and $A$ dependence
of distortions in all nuclei with mass $A \geq$ 4.

In this paper we present new evidence for such
universality,  found in the analysis
of recent data collected from the DIS of muons on nuclei
by the NMC (CERN)~\cite{ama95,arn95}
 and E665 (Fermilab)~\cite{ad95} collaborations.
This data brings to 14 the number of nuclei studied in the
DIS of muons and electrons, which offers a good
opportunity for studying the $A$ dependence of
distortions in the structure function in nuclei
from $^4$He to $^{207}$Pb.\\
%
%

 Below, we consider
structure function distortions as independent of
the  $Q^2$ at which $r^A(x)$
is measured. This is justified by  conclusions about the
$Q^2$ independence of $r^A$ in the range  0.2 GeV$^2$
$< Q^2 <$ 200 GeV$^2$ (c.f. Refs.~\cite{ama95}--\cite{bari}).

  In Ref.~\cite{sm94} it was found that the $x$ dependence of
$r^A(x)$ can be factorized into three parts
in the region  0.003 $< x <$ 0.7, in
accordance with the differences in the $r^A(x)$ behaviour
found in the three intervals of the considered range ---
namely  the (1) shadowing,
(2) anti-shadowing and (3) EMC effect regions:
\begin{equation}
r^A(x) \equiv F_2^A(x) / F_2^D(x) ~=~ x^{m_1} (1+m_2) (1- m_3 x) .
\label{smir}
\end{equation}
\noindent The parameters $m_{i}$, $i$= 1 -- 3,
 can be treated as  the
distortion magnitude of the nucleon structure function
introduced for each interval. There are two physical reasons for
parametrizing $r^A(x)$ in the form of Eq. (\ref{smir}).
First, as was shown in Ref.~\cite{marti}, the nucleon structure
function behaves as $F_2(x) \sim x^{- \lambda}$
in the range of small $x$, which is motivated by BFKL dynamics.
Hence,  combinations such as  $F_2^A(x) / F_2^D(x)$ should obey
a power law as well. Second, the parameters $m_2$ and $m_3$
enter Eq. (\ref{smir}) in a manner similar to the suggestion
of Ref.~\cite{rep160}, whereby local nuclear density is related
 to the deviation of $r^A(x)$ from unity in the range $x >$ 0.3.

As is shown in Ref.~\cite{sm94}, the use of Eq. (\ref{smir}) is
justified in the range 0.5 $< Q^2 <$ 200 GeV$^2$. ~Nuclear
shadowing~ is then~ described by one term only,~ since for\\
 $x \ll$ 1, Eq. (\ref{smir}) reduces to the relation
\begin{equation}
r^A(x) ~=~ C x^{\alpha} ~.
\label{shad}
\end{equation}

\noindent A similar rate of increase in shadowing with a decrease
in $x$ was expected at high $Q^2$,
due to the gluon fusion considered in
Refs.~\cite{baron}. The results of Ref.~\cite{sm94} thus
indicate that the gluon fusion mechanism persists for
$Q^2$ as low as $\sim$0.5 GeV$^2$. The NMC and E665 data
obtained in the range below $x$ = 0.003 do not deviate from
Eq. (\ref{shad}) until
$x$ = 9$\cdot 10^{-4}$, $Q^2 \approx$ 0.2 GeV$^2$.
At the lower values of $x$, which correspond in the
kinematics of NMC and E665 to lower $Q^2$, the data indicate
 (c.f. Refs.~\cite{arn95,ad95,xe92,pa94} )  a smooth
transition to the values of photoabsorption cross-section ratios,
and thus cannot reflect the distortions of parton distributions
by the nuclear medium. Therefore, we considered
the data in the range $Q^2 >$ 0.3 GeV$^2$, which excludes
the transition region  $x<$ 0.001.

The parameters $m_i$ were determined by fitting $r^A(x)$,
measured on seven nuclear targets --- He~\cite{ama95,gomez},
Li~\cite{arn95}, C~\cite{arn95,gomez}, Ca~\cite{ama95,gomez},
Xe~\cite{xe92}, Cu~\cite{copper} and Pb~\cite{ad95} ---
 with Eq. (\ref{smir}).
We used in the fit the total experimental error determined
by adding statistical and systematic errors at each point
in quadrature. For each of seven nuclei,  good agreement
($\chi ^2 /$d.o.f.$ \leq$ 1) with Eq. (\ref{smir}) was found,
thus proving  that the characteristic pattern   of the
structure function modifications,  well described
for the helium nucleus by Eq. (\ref{smir}),
remains unchanged for heavier nuclei. We consider this
a manifestation of the universality of the $x$ dependence
of the distortions of the free-nucleon structure function
in a nuclear environment.\\

The results of the fit are shown in Fig. 1. The obtained
parameters $m_i$, which represent the distortion magnitudes,
increase from their minimum value $m_i$(He) at $A$ = 4
to $m_i(A)$ $\approx$ 3$m_i$(He) for $A>$ 40,
indicating that  distortions in heavy nuclei are
independent of the size of the nucleus.
Previously, saturation of distortions was observed
in the EMC effect region from the analysis of data
available in the range 4 $\leq A \leq$ 197~\cite{sm94}.

The  parameters $m_i$  vary similarly with  $A$
in all three intervals in which the distortions
were depicted. This similarity was first observed in
Ref.~\cite{sm94} on a smaller sample of data.
The points in Fig. 1 are approximated by the following
equation:
\begin{equation}
  m_i (A) = N_i \Biggl( 1 - \frac{A_{\rm S}}{A} \Biggr)  ~.\\
\label{smbar}
\end{equation}
This coincides, except  for the normalization parameter $N_i$,
with the factor $\delta (A)$ suggested in Ref.~\cite{barsh}
for explaining the $A$ dependence of the EMC effect:
\begin{equation}
\delta(A) = N \Biggl( 1 - \frac{A_{\rm S}}{A} \Biggr)=
N \Biggl( 1- \frac{1}{A^{1/3}} -\frac{1.145}{A^{2/3}}+\frac{0.93}{A}+
 \frac{0.88}{A^{4/3}} - \frac{0.59}{A^{5/3}} \Biggr)~  ,
\label{bars}
\end{equation}
\noindent where
the number of nucleons $A_{\rm S}$ at the nuclear surface was
obtained using a Woods--Saxon potential with parameters taken from
Ref.~\cite{bohr}:
\begin{equation}
A_{\rm S} = 4 \pi \rho_0 \int \limits_{r_0(A)}^{\infty} dr r^2
{ {1} \over {1 + e^{[r-r_0(A)]} / a} }~,
\label{woods}
\end{equation}

\noindent where $\rho_0$ = 0.17 fm$^{-3}$
is the central nuclear density,
 $r_0(A)$ = (1.12$A^{1/3}$ -- 0.86$A^{-1/3}$) fm is a nuclear radius,
and $a$ = 0.54 fm is the nuclear surface diffusion coefficient.

Thus the three lines in Fig. 1, a, b and c, differ only in the
normalization factor $N_i$, which was found to be
$N_1$ = 0.130 $\pm$ 0.004
for the shadowing region, $N_2$ = 0.456 $\pm$ 0.017 for the
anti-shadowing region and $N_3$ = 0.773 $\pm$ 0.020
for the  EMC effect region.
As shown in the comparison of $m_i$ with Eq. (\ref{smbar}),
the $A$ dependence of the distortion magnitudes is
consistent with that
 defined by Eqs. (\ref{smbar}) and (\ref{woods}).
In other  words, our results give evidence for
a {\em universal} $A$ dependence of the distortion magnitudes
$m_i$ of the nucleon structure function in
all three regions.  This universality  can be expressed
in terms of the relative  distortions, measured in
nuclei  $A_1$ and $A_2$ with the following relation:
\begin{equation}
 {m_1(A_2) \over m_1(A_1) } ~=~
{m_2(A_2) \over m_2(A_1) } ~=~ {m_3(A_2) \over m_3(A_1) } ~.
\label{simil}
\end{equation}

\begin{figure}[p]
\begin{center}
\mbox{\epsfxsize=1.08\hsize\epsffile{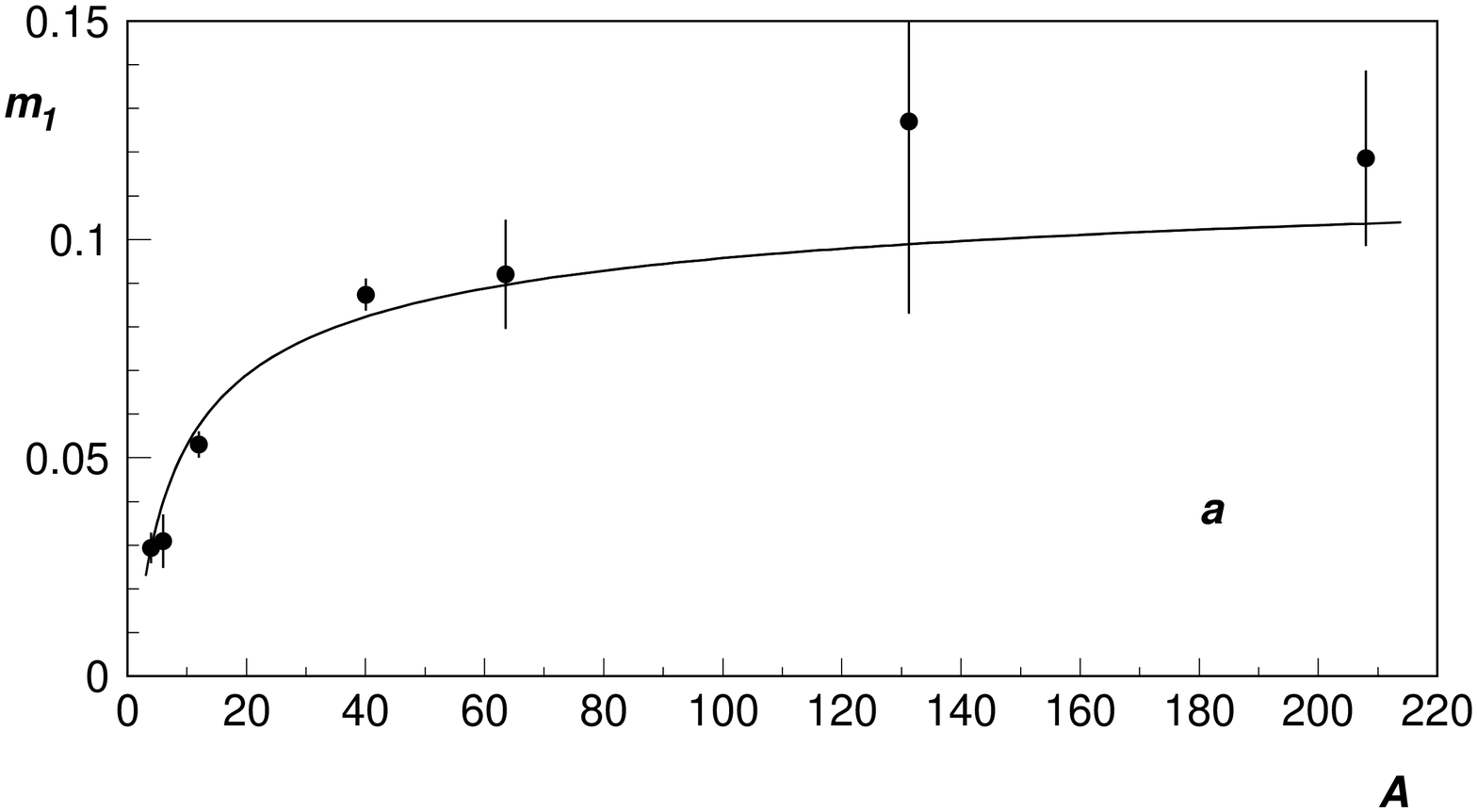}}
\end{center}
\vspace{-0.5cm}
\begin{minipage}[t]{0.485\linewidth}
\begin{center}
\mbox{\epsfysize=\hsize\epsffile{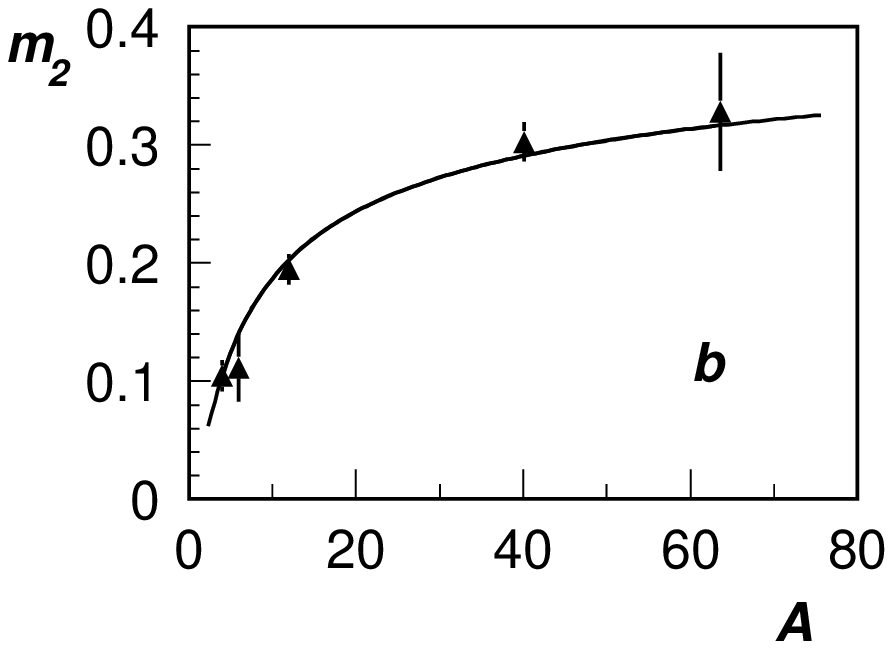}}
\end{center}
\end{minipage}
\hfill
\begin{minipage}[t]{0.485\linewidth}
\begin{center}
\mbox{\epsfysize=\hsize\epsffile{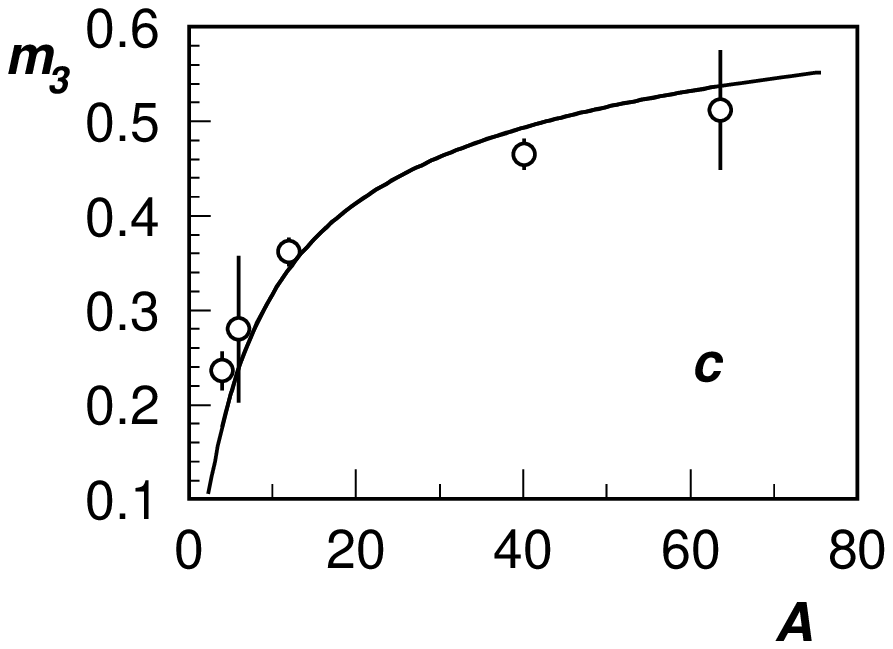}}
\end{center}
\end{minipage}
\caption{
 The parameters $m_i$, $i$= 1 -- 3, which define the magnitude
 of distortions of the nucleon structure function in a nuclear
 environment as a function of atomic mass $A$,
 determined in the  regions of nuclear shadowing (a),
 anti-shadowing (b) and the EMC effect (c). Full lines show
 a variation in nuclear density given by the Woods--Saxon
 potential, with parameters fixed from the data on elastic
 electron--nucleus scattering. The three lines differ only in
 the normalization  found from the fit to
 $m_i$.}

\end{figure}

One can as well define the value of structure
function distortion in units of that measured in the helium nucleus,
{}~$s_{\rm h}$ = $m_i(A)$ / $m_i$(He).
{}~ By definition, \\
$s_{\rm h}$ = 1 for $A$ = 4, and,
as follows from the obtained numerical values of $m_i$,
$s_{\rm h}$ increases with $A$ to $\sim$ 3 for heavy nuclei,
independent of $x$.\\

The universality of the $x$ dependence of the nucleon structure
function distortions implies that the
positions of the three cross-over points $x_i$, $i$ = 1 -- 3,
in which $r^A(x)$ = 1, are $A$-independent if $A \geq$ 4.
Until recently, large
experimental errors did not allow verification of theoretical
predictions on the position and $A$ dependence of $x_i$,
 discussed in  a number of publications
 (c.f. Refs.~\cite{qiu87}--\cite{zhu90}).
The situation has not improved for  $x_{\rm III}$ ($\sim$ 0.8),
and one needs both higher statistical accuracy
in  $r^A(x)$ and a larger number of nuclei to establish whether
$x_{\rm III}$ is  indeed $A$-independent. On the other hand,
the data on $r^A(x)$ currently available in the EMC effect
region  made it possible to establish
that, within experimental errors, the coordinate of the second
cross-over point does not depend upon $A$  in the range
4 $ \leq A \leq $ 197  and equals
 $x_{\rm II} $= 0.273 $\pm$ 0.010~\cite{sm94}.\\

 We find  $x_{\rm I}$ as an
intersection  point of a straight line $r^A(x)$ = 1, with
$r^A(x) $ given by Eq. (\ref{shad}).
The parameters $C$ and $ \alpha$ have been found by
fitting DIS data in the range  0.001 $<x<$ 0.08
on He, Li, C and Ca by NMC~\cite{ama95,arn95}, on Cu
by EMC~\cite{copper}, and on  Xe~\cite{xe92} and Pb~\cite{ad95} by
E665.

Agreement between the data obtained on the same nuclear target from
two different experiments is an absolutely necessary condition for
including data from the two experiments in a study of the $A$
dependence. As shown in Ref.~\cite{ad95}, the data on $r^{\rm C}(x)$
and  $r^{\rm Ca}(x)$ from NMC and E665 are not consistent with
each other in the range $x<0.1$  and thus can not be combined
for the analysis.
As in the case of the $A$ dependence of $m_i$, one would expect that
the $A$ dependence of $x_{\rm I}$ shows itself in the range $A <$ 40.
Consequently, in order to minimize systematic errors  we have kept
in the analysis the data  on C and Ca nuclei from the NMC, which
complement the data on He and Li from the same collaboration.
At the same time we have included the data on the lead nucleus,
 collected by E665 only. For  consistency's
sake we use the  $r^{\rm Pb}(x)$ from Ref.~\cite{ad95} which
was obtained using the NMC procedure of radiative corrections.

The  values $x_{\rm I}$ obtained as a function of $A$
are plotted in Fig.~2.
Similar to the behaviour observed earlier for $x_{\rm II} $,
 within experimental errors the results are consistent
with  $x_{\rm I}$ = const ($\chi^2$/d.o.f. = 6.1/7)
and correspond to $x_{\rm I}$ = 0.0615 $\pm$ 0.0024 .

 \begin{figure}[t]

\begin{center}
  \mbox{\epsfxsize=0.9\hsize\epsffile{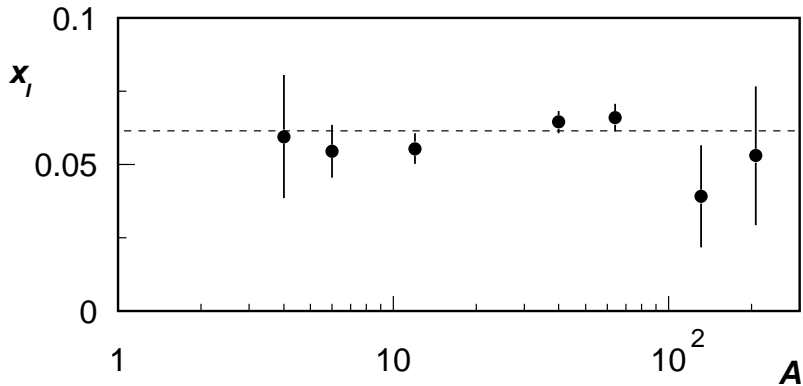}}
\end{center}
\caption{ The coordinate of the first cross-over
point $x_{\rm I}$ as a function of atomic mass $A$. The average
value $\bar x_{\rm I}$ = 0.0615 is shown with a dashed line.}

\vfill
\end{figure}

The $A$-{\em independence} of $x_{\rm I}$
demonstrated by the present analysis provides a clue for
better understanding of the shadowing mechanism.
As follows from the results shown in Fig. 2, the idea of the
$A$-dependent  $x_{\rm I} $, widely exploited in
models~\cite{berg88}--\cite{zhu90},  has to be discarded.
Going back to the first discussion of the shadowing in the parton
model~\cite{nik75}, a number of papers relate the onset of
the deviation of $r^A(x)$ from unity to the size of the
region where the partons belonging to two or more
neighbouring nucleons can be
localized~\cite{berg88}--\cite{zhu90}. The position of $x_{\rm I}$
is related  in Ref.~\cite{berg88} to the size
of a nucleon, $R_{\rm N}$, and the number of overlapping nucleons $n$:
\begin{equation}
x_{\rm I} = { 1 \over {(n-1) 2 R_{\rm N} m_{\rm N}}} ~,
\label{berg}
\end{equation}
\noindent where $m_{\rm N}$ is the nucleon mass.
Taking  our result for $x_{\rm I}$ and also the most precise value of
the proton root-mean-square radius
 $< R_{\rm E}^2 >_p^{1/2}$= 0.862 $\pm$ 0.012 fm, obtained
from the analysis of the data on elastic electron--proton
scattering~\cite{prot80}, we find that $n$ = 2.98 $\pm$ 0.08.\\

The suggestion of ref.~\cite{barsh} to use the nuclear
surface-to-volume ratio to explain the modification
of the nucleon structure function in the EMC effect region
has also been  explicitly considered in Refs.~\cite{date,sick}.
Our observation of the $A$ dependence of $m_i$ means that
 the nucleon structure is not modified
if the nucleon belongs to the nuclear surface, not only in the
EMC effect region, but also in the regions of nuclear
shadowing and anti-shadowing.

The parameters describing  nuclear structure in Eq. (\ref{woods})
have been determined from the elastic scattering of electrons
on nuclei (c.f. Ref.~\cite{hof}).  The same parameters allow one to
reproduce with Eq. (\ref{smbar}) the $A$  dependence of  $m_i$,
obtained from experiments with momentum transfers of three orders of
magnitude higher.  Thus, the results shown in Fig. 1
demonstrate remarkable consistency between experimental
studies of nuclear structure from {\em deep-inelastic} and
{\em elastic} scattering of leptons off nuclei.

The role of the nuclear surface-to-volume ratio in the
observed modifications of the nucleon structure function
can also be studied by means of the comparison of $F_2^{A_1}(x)$
 and  $F_2^{A_2}(x)$, when $A_1$, $A_2 \geq$ 4.
The results of such measurements are expected soon from
the NMC Collaboration~\cite{mid}.
 When $F_2^{A}(x)$ is obtained
in the DIS regime in the shadowing region, the ratio of the
structure functions is described by
Eq. (\ref{shad}), where the parameter $\alpha (A_1 /A_2)$
is related to distortions $m_1(A)$ in a trivial way:
\begin{equation}
\alpha (A_1 / A_2) ~=~ m_1(A_1) ~-~ m_1(A_2)~.
\label{a1a2}
\end{equation}

Obviously, $ \alpha (A_1 / A_2)$ can also be calculated from
the data on the EMC effect by using Eq. (\ref{smbar}) and
the normalization parameters $N_i$ found by our analysis.
The similarity in the $A$ dependence  of the $m_i$ justifies
the use of Eq. (\ref{simil}) to relate  $ \alpha (A_1 / A_2)$
to $m_3(A)$, even if deviations from Eq. (\ref{smbar})
are found.

{}From the universality of the $x$ dependence of the distortions
 we expect that
the coordinate of the first cross-over point determined from
the ratios  $F_2^{A_1}(x)$/$F_2^{A_2}(x)$ is $A$-independent
and coincides with that determined by the present analysis.\\

%
%
Perturbative QCD provides a natural framework for the
calculation of the modification to the structure function arising
from the fusion of quarks, anti-quarks
 and gluons~\cite{qiu87,muel,clo89}.
As has been shown in Ref.~\cite{clo89} QCD (together with effects of
$Q^2$ rescaling) is capable of describing the
modifications to $F_2(x,Q^2)$ not only in the nuclear shadowing
region, where it proved to be very successful, but also in the entire
$x$ range. There remains, however, the problem of the role
of two-, three- and four-parton fusion mechanisms in the QCD
calculations.
Judging from the agreement between the data on $r^{\rm He}(x)$ and
calculations which assume either a two-~\cite{baron} or
three-~\cite{barsh}  gluon fusion mechanism,
one cannot give preference to either of the two approaches. New
insight into this problem is provided by our results on $x_{\rm I}$,
which should be considered as an argument in favour of
contribution of the recombinations of gluons from three different
 nucleons.

Further improvement of the theoretical description
of distortions in a free-nucleon structure function
is hardly possible  until the mechanism responsible for the
universality of the $x$ and $A$ dependence of the EMC effect
is fully understood.

We suggest that modifications to the
parton distributions of the nucleon bound in a nucleus
 evolve as a function of atomic mass $A$
in two stages. In the first stage, the distributions of
partons belonging to the
lightest nuclei, 2 $< A \leq$ 4, are modified drastically
 compared to those of a free nucleon, thus distorting
the  structure function $F_2(x)$.
These distortions, which can be observed in a $^4$He
nucleus as  a characteristic oscillation of $r^A$
around the line  $r^A$ = 1,  remain frozen in shape
in the second stage of distortions, which occur in nuclei with
mass  $A >$ 4.
In contrast to the first stage, in the second
 there is no restructuring of parton distributions,
 which can change the shape
of the oscillation described by Eq. (\ref{smir}). Instead, the
distortions increase in magnitude throughout the
entire $x$ range, following the functional form (\ref{smbar}).

There  are evidently two different mechanisms behind this
picture, which we denote as {\em hard} or {\em soft}
distortions, depending on whether $A \leq$ 4 or $A >$ 4.
Quantitatively, this can be expressed with the parameter $s_{\rm h}$,
which rapidly changes in the range of hard distortions,
from 0 to 1 ($\Delta A$ = 2), and only slowly in the range of
soft distortions, from 1 to $\sim3$ ($\Delta A \approx$ 200).
A particular case of the hard distortion mechanism, which works at
$A$ = 4,  has been
considered in Refs.~\cite{fk83,kon84}, in which EMC effect was
explained by the 12-quark  structure of nuclei.

In terms of the two-mechanism  model, the experimental
observations can be interpreted as follows: a) the
positions of the three cross-over points are determined by hard distortions,
and b) $x_i$ are $A$-independent in the range of soft
distortions. In other words, hard distortions are saturated at $A$ = 4,
which can be understood if modifications of parton distributions in the nuclear
environment are closely related to short-range nuclear forces.
In this picture  $x_{\rm III}$ should be different
when it is obtained in  $^3$He and $^4$He nuclei. Before such data are
available one can not  exclude the possibility that the saturation
is  reached at $A$ = 3.\\

In summary,
we have shown that the recent data on the DIS of electrons
and muons off nuclei bring new evidence for the universality of
the $x$ and $A$ dependence of distortions of a free-nucleon
structure function, $F_2(x)$, by a nuclear medium, when $A \geq$ 4.
Such  universality and, in particular, the
evidence for the $A$-independence of  $x_{\rm I}$, imply
that hard distortions of parton distributions
are saturated at $A$ = 4 (or even at $A$ = 3) and that
the observed differences between the DIS
cross-sections for nuclei with masses $A_1$, $A_2 \geq$ 4
are due to soft distortions.
The latter are similar in the shadowing, anti-shadowing and
EMC effect regions, and vary from 1 in $^4$He to $\sim$3 in
$^{207}$Pb. They  can
be well understood as a nuclear density effect if the
surface nucleons are excluded from consideration.

 It has been found that nuclear
shadowing begins at $x_{\rm I}$ = 0.0615 $\pm$ 0.0024,
which is consistent with models that relate $x_{\rm I}$
to a picture of  the recombination of partons from three
different nucleons.

The problem of describing modifications of  $F_2(x)$
 in a nuclear medium can thus be reduced to
the derivation of $F_2^{\rm He}(x)$.
We see further progress in this field in experimental studies
of  hard distortions of the structure function
 in the $^3$He nucleus, and
also in the search for possible deviations from the
$A$ dependence of $r^A(x)$ defined by the surface-to-volume
ratio (e.g. saturation of soft distortions in heavy nuclei).\\

{\large Acknowledgements}

The author would like to acknowledge helpful discussions with
A.M. Baldin, S.B. Gerasimov, A.V. Efremov, L.L. Frankfurt,
V.K. Luk'yanov, I.A. Savin and   M.V. Tokarev.
He is also grateful to E. Gabathuler,
D. Kharzeev, B.Z. Kopeliovich, G. van Middelkoop,
U.-G. Meissner,  N.N. Nikolaev  and V.D. Toneev
for useful remarks.\\



\end{document}